\documentclass[aps,prl,twocolumn,groupedaddress]{revtex4-1}

\usepackage{graphicx,epstopdf,siunitx,braket,xspace,amssymb,dsfont,bm}

\newcommand{\qS}[0]{\ensuremath{\Ket{S}}\xspace}
\newcommand{\qT}[0]{\ensuremath{\Ket{T_0}}\xspace}
\newcommand{\jj}[0]{\ensuremath{J}\xspace}
\newcommand{\rfig}[1]{Fig.\,\ref{#1}}

\newcommand{\rtab}[1]{Tab.\,\ref{#1}}
\newcommand{\mc}[1]{\ensuremath{\mathcal{#1}}}
\newcommand{\mr}[1]{\ensuremath{\mathrm{#1}}}
\newcommand{\je}[0]{\ensuremath{J(\epsilon)}\xspace}
\newcommand{\dbz}[0]{\ensuremath{\mr{\Delta}B_z}\xspace}
\newcommand{\eps}[0]{\ensuremath{\epsilon}\xspace}
\newcommand{\F}[0]{\ensuremath{\mc{F}}\xspace}
\newcommand{\IF}[0]{\ensuremath{\mc{I}}\xspace}
\newcommand{\qp}[0]{\ensuremath{\mc{E}}\xspace}
\newcommand{\ndbz}[0]{\ensuremath{N_{\dbz}}\xspace}
\newcommand{\bvec}[1]{\ensuremath{\mathbf{#1}}}

\newcommand{\bveceps}[0]{\ensuremath{\bm{\epsilon}}}
\newcommand{\pbveceps}[0]{\ensuremath{\bm{\epsilon}}'}

\newcommand{\figlabel}[1]{(#1)}

\newcommand{\nseg}[0]{\ensuremath{N_{\mathrm{seg}}}\xspace}

\begin{document}


\title{High-Fidelity Single-Qubit Gates for Two-Electron Spin Qubits in GaAs}


\author{Pascal Cerfontaine\textsuperscript{1}}
\email[]{pascal.cerfontaine@rwth-aachen.de}
\author{Tim Botzem\textsuperscript{1}}
\author{David P. DiVincenzo\textsuperscript{1, 2}}
\author{Hendrik Bluhm\textsuperscript{1}}
\affiliation{\textsuperscript{1}JARA-Institute for Quantum Information, RWTH Aachen University, D-52074 Aachen, Germany}
\affiliation{\textsuperscript{2}Peter Gr\"unberg Institute: Theoretical Nanoelectronics, Research Center J\"ulich, D-52425 J\"ulich, Germany}


\date{October 7, 2014}

\begin{abstract}
Single-qubit operations on singlet-triplet qubits in GaAs double quantum dots have not yet reached the fidelities required for fault-tolerant quantum information processing. Considering experimentally important constraints and using measured noise spectra, we numerically minimize the effect of decoherence (including high-frequency 1/f-like noise) and show theoretically that quantum gates with fidelities higher than 99.9\% are achievable. We also present a self-consistent tuning protocol which should allow the elimination of individual systematic gate errors directly in an experiment.
\end{abstract}

\pacs{}

\maketitle

One well-established possibility to realize a qubit with electron spins in a semiconductor is to use the $m_s=0$ spin singlet and triplet states of two electrons as computational basis states \cite{Levy2002}. In contrast to single electron spins this encoding allows for all-electrical qubit control. Very long coherence times of up to \SI{200}{\micro s} \cite{Bluhm2010a}, all aspects of single-qubit operation (e.g. initialization \cite{Petta2005} and single-shot readout \cite{Barthel2009}) and a first two-qubit gate \cite{Shulman2012} have been demonstrated experimentally for such singlet-triplet (ST) qubits in GaAs quantum dots. Universal single-qubit control was also shown \cite{Foletti2009} but subject to large uncharacterized errors. Limiting control error rates to $\sim 10^{-3}$ is a crucial requirement for fault tolerant quantum computing with quantum error correction (QEC) \cite{Fowler2009, Raussendorf2007, Knill1998a}. Estimates based on coherence time measurements \cite{Bluhm2010a, Dial2013} indicate that very high gate fidelities should be possible for GaAs-based two-electron spin qubits. However, nonlinearities in the electric control and experimental constraints make the direct application of control methods such as Rabi driving difficult.

Previous theoretical work has shown how universal control on the single- and two-qubit level can be achieved in the face of limited dynamic control range \cite{Hanson2007}. Additionally, gating sequences which are insensitive to slow (quasistatic) control fluctuations have been proposed for this qubit system \cite{Grace2012, Kestner2013, Wang2014, Khodjasteh2012}. While these proposals provide very useful conceptual guidance, a direct implementation will be impeded by experimental constraints such as finite pulse rise times and sampling rate of voltage pulses. Likewise, decoherence effects caused by charge noise \cite{Dial2013} and nuclear spin fluctuations \cite{Reilly2008} have a significant effect. 

In this letter we use numerical pulse optimization to address systematic inaccuracies and decoherence. Pulse optimization is common in NMR \cite{Khaneja2005} and is also receiving increasing attention in quantum information \cite{Bylander2009, Grace2012, Grace2013, Jelezko2014, Egger2013, Schutjens2013, Egger2014, Kabytayev2014}. In contrast to these previous approaches, our optimization is specifically tailored to the ST-qubit system and includes not only the relevant physical effects but also the most important hardware constraints and the effect of high-frequency 1/f-like noise. We use experimentally determined parameters and noise spectra \cite{Dial2013, Bluhm2010, Reilly2008} to compute expected gate fidelities \F and find implementations with no systematic errors and optimized robustness to both slow and fast noise. With this approach we show that $\pi$- and $\pi/2$-gates around orthogonal axes with \F exceeding $99.9\%$ can be achieved.

Reaching these high fidelities experimentally is complicated by the difficulty of characterizing experimental parameters to a sufficient degree of accuracy. Therefore, we propose a self-consistent calibration routine which iteratively tunes pulse sequences using feedback from the experiment \cite{Dobrovitski2010a}. We benchmark this routine via simulations and show that it allows the elimination of systematic errors that arise when the numerical pulses are applied on the experiment. This justifies neglecting less relevant systematic effects in the pulse optimization.

\begin{figure}
	\centering
	\includegraphics[width=6.8cm]{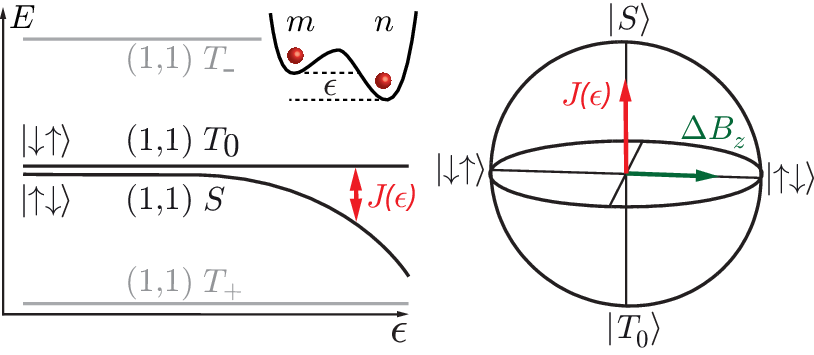}
	\caption{(color online) Left: Energy diagram of the computational subspace (black) as a function of \eps. The transfer function \je is nonlinear and modeled as $\je = J_0 \exp{(\eps/\eps_0)}$. Right: Bloch sphere convention.}
	\label{fig:bloch}	
\end{figure}

In the gate-defined quantum dots considered here, the double quantum dot used to hold the two electrons is formed from a two-dimensional electron gas by applying voltages to surface gates on a GaAs/AlGaAs-heterostructure. The potential difference $\epsilon$ between the two dots changes the charge configuration (m, n), where m (n) is the number of electrons in the left (right) dot (\rfig{fig:bloch}). Computation is performed in (1,1) using the subspace spanned by the spin singlet state $\qS = (\ket{\uparrow\downarrow}-\ket{\downarrow\uparrow})/\sqrt{2} = \ket{0}$ and the $m_s = 0$ triplet state $\qT = (\ket{\uparrow\downarrow}+\ket{\downarrow\uparrow})/\sqrt{2} = \ket{1}$. The $m_s = \pm 1$ (1,1) triplets are Zeeman split by an external magnetic field of typically more than \SI{100}{mT} (\rfig{fig:bloch}) \cite{Petta2005}. In the following we focus only on the computational subspace since the leakage probability to states with different charge and spin configurations is lower than \SI{5e-5} for the presented pulse sequences, verified numerically using a seven-level Hamiltonian with spin-orbit interaction \cite{supp}.

Since only \qS can tunnel from (1,1) to (0,2), the spin state can be read out by spin to charge conversion \cite{Petta2005}. The tunnel coupling also leads to an \eps-dependent exchange energy \jj between \qS and \qT (\rfig{fig:bloch}). Additionally, each electron spin couples via the hyperfine interaction to a different nuclear spin environment in each dot. This interaction can be described by a magnetic `Overhauser' field gradient \dbz between the dots, and creates an energy difference between $\ket{\downarrow\uparrow}$ and $\ket{\uparrow\downarrow}$ \cite{Petta2005}. The Hamiltonian can then be written in the $\{\qS, \qT\} $ basis as $H = \frac{\hbar \je}{2} \sigma_z + \frac{\hbar \dbz}{2} \sigma_x$ with Pauli matrices $\sigma_i$ and \dbz in units of angular frequency.

In typical experiments, arbitrary waveform generators (AWGs) are used to produce pulses $\eps(t)$ which control \je. Since \dbz can be set to any desired constant value by dynamic nuclear polarization \cite{Bluhm2010}, it is possible to realize arbitrary single-qubit target gates $U_t$ \cite{Foletti2009}. Systematic deviations from $U_t$ arise mainly from finite rise times of the voltage pulses and a nonlinear and imperfectly characterized transfer function \je. In addition, two sources of noise lead to significant decoherence. While fluctuations in \dbz are much slower than typical gate operation times ($\sim \SI{10}{ns}$), charge noise affects \eps also on much shorter timescales \cite{Reilly2008, Dial2013}. 

All the above effects are accounted for in our numerical simulations. We use a phenomenological model $J(\eps(t)) = J_0 \exp(\epsilon(t)/\epsilon_0)$ determined from fits to experimental data \cite{Dial2013}. The fixed sample rate of AWGs is modeled with rectangular pulses in \eps with a fixed sample duration. This results in amplitude-only control in each of \nseg pulses $\eps_j, j = 1 \ldots \nseg$, with bounds $\eps_{\min} \le \eps \le \eps_{\max}$. Furthermore, we model finite rise times, due to AWG limitations, the skin effect in coaxial cables and stray capacitances, as exponential with a time constant $\tau_{\mr{rise}} \sim \SI{1}{ns}$. In addition, we enforce a waiting period of $4 \tau_{\mr{rise}}$ at the end of each gate to give \eps time to decay to a predefined baseline $\eps_{\min}$. This allows for straightforward concatenation of different gate sequences since transients from previous gates are minimized. For the use in a quantum processor it may be convenient for different gates to have the same duration $T$, providing the quantum system with a clock rate as in classical computers. Likewise, it is attractive to be able to leave the current qubit state unchanged over one or several clock cycle periods. This is most easily done by \dbz-rotations with $\sqrt{\dbz^2 + J(\eps_{\min})^2} \, T = 2 \pi \ndbz$, where $\ndbz \in \mathds{N}$ gives the number of \dbz-rotations and the exchange splitting is kept constant at $J(\eps_{\min}) \ll \dbz$.

\begin{figure}
	\centering
	\begin{picture}(8.6,5.9)
		\put(0.315,0){\includegraphics[width=8.6cm,trim=0cm 0cm 0cm 0cm,clip=true]{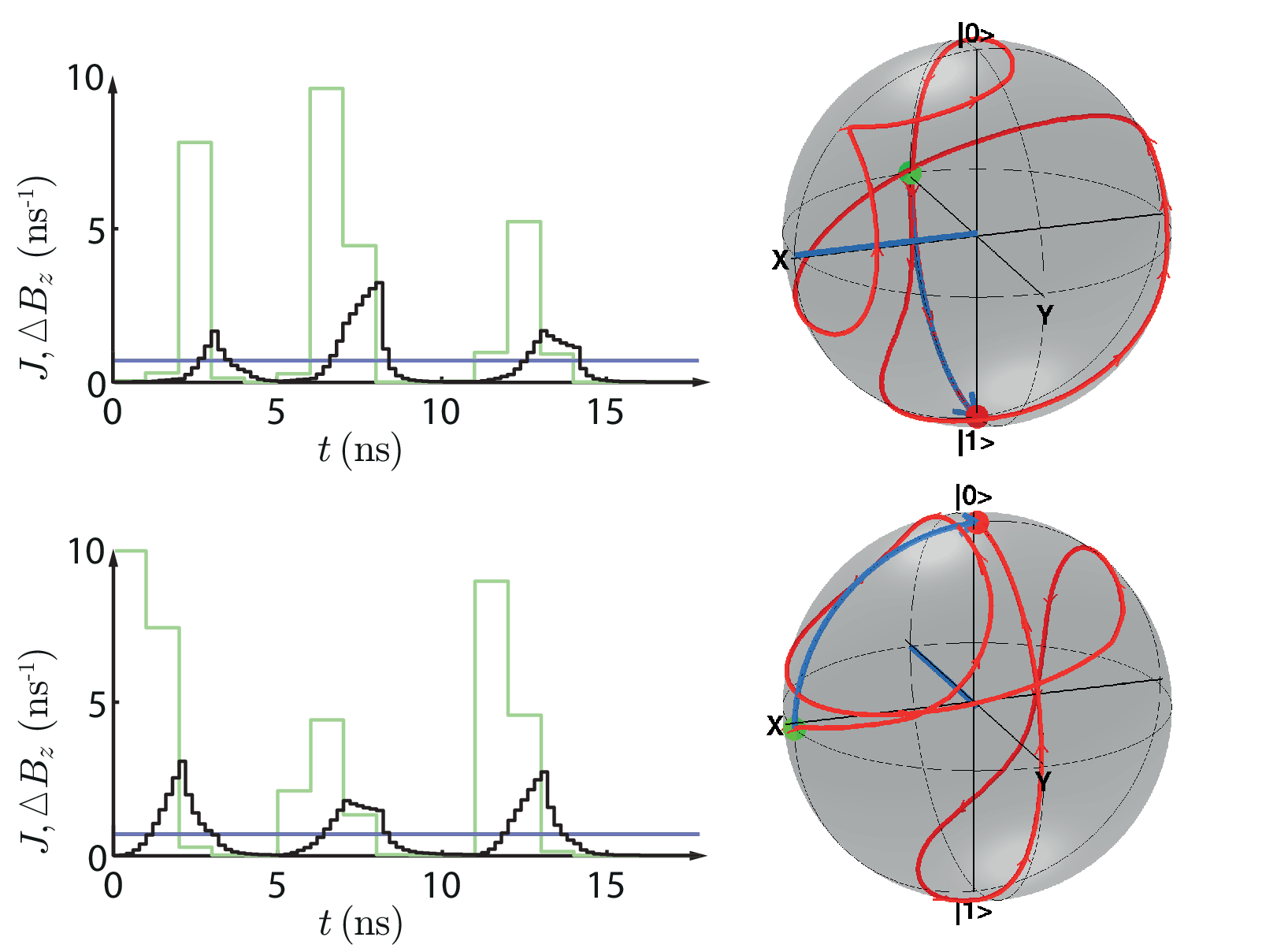}}		
		\put(0.1,5.8){\figlabel{a}}	
		\put(0.1,2.65){\figlabel{b}}
  \end{picture}
	\caption{(color online) \figlabel{a} $\pi/2_x$-gate with $\IF = 1 - \F = \SI{1.5e-3}{}$ \figlabel{b} $\pi/2_y$-gate with $\IF = \SI{1.6e-3}{}$. Rectangular \jj-pulses are shown in green, black lines show $J(t)$ when accounting for finite rise times and \dbz is shown in blue. The corresponding Bloch sphere trajectories for both pulses are plotted for selected initial states (green dot).}
	\label{fig:gates}	
\end{figure}

Taken together, these constraints result in a discrete set of acceptable values for \dbz and in pulse shapes $J(t)$ as in \rfig{fig:gates}. Since the calibration routine discussed below can remove relatively large systematic errors, it is sufficient to qualitatively describe the system in the simulations and correct quantitative inaccuracies by using experimental feedback.

In simulations we approximate explicitly time-dependent Hamiltonians $H(J(\eps(t)),\dbz)$ as piecewise constant. For appropriate discretization this simplification incurs negligible errors but makes the calculation of $U(t,t_0) = \mc{T} \exp\left(-\frac{i}{\hbar}\int_{t_{0}}^{t} H(t')\ dt'\right)$ straightforward. We use the average gate fidelity \F \cite{Bowdrey2002, Nielsen2002} between $U_t$ and a quantum process \qp as an objective function in numerical pulse optimization. To compute the effect of quasistatic noise we sample discretely from a Gaussian distribution, for fast noise we use a first-order perturbative approach \cite{Green2012} which allows for swift evaluation of the infidelity $\IF = 1-\F$, suitable for numerical optimization.

The offset $\delta \dbz$ from a stabilized \dbz varies slowly ($\gtrsim \SI{0.1}{s}$) compared to gate operations with a measured standard deviation $\sigma_{\dbz } \approx \SI{0.5}{mT}$ \cite{Reilly2008, Bluhm2010}. For low- and high-frequency charge noise we use recent measurements of the standard deviation and spectral noise density, given as $\sigma_{\eps} = \SI{8}{\micro V}$ and $S_{\eps}(f) = \SI{8e-16}{} \frac{\mr{V}^2}{\mr{Hz}} \left( \frac{\SI{}{Hz}}{f}\right)^{0.7}$ from \SI{50}{kHz} to \SI{1}{MHz} \cite{Dial2013}. We conservatively extend the spectrum as white above \SI{1}{MHz} until \SI{3}{GHz}, using $S_{\eps}(\SI{1}{MHz})$. Choosing the upper cutoff higher than \SI{3}{GHz} does not influence the calculated impact of \eps-noise on gate performance. A Taylor expansion of \je yields $J(\eps(t) + \delta \eps(t)) \approx J(t) \cdot (1 + \delta \eps(t) / \eps_0)$. 

\begin{figure}
	\centering
	\begin{picture}(7.5,4.2)
		\put(0,0){\includegraphics[width=7.5cm]{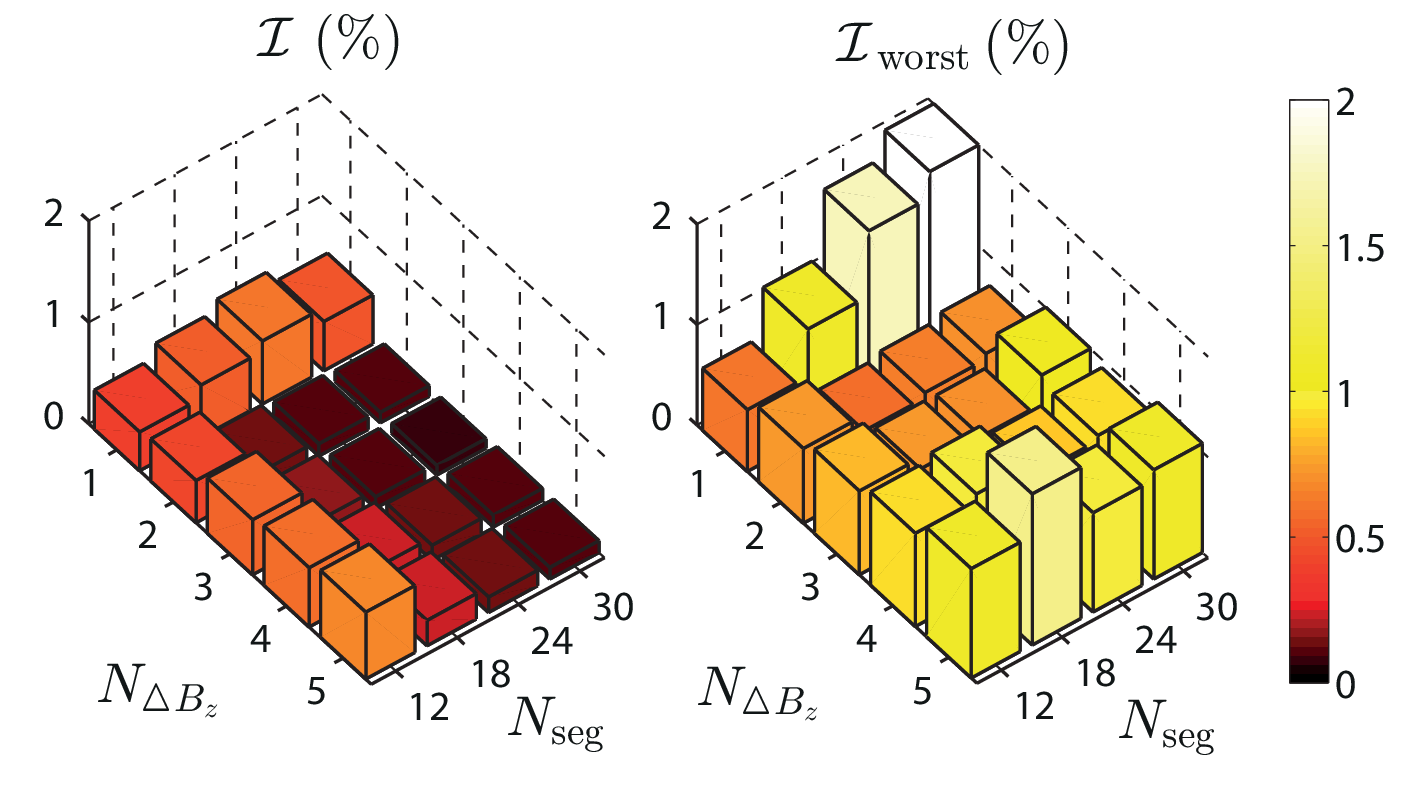}}
		\put(0.25,3.9){\figlabel{a}}
		\put(3.45,3.9){\figlabel{b}}
  \end{picture}
	\caption{(color online) Infidelities of $\pi/2_x$-gates \figlabel{a} Best solutions for different number of pure \dbz-rotations and number of \eps-pulses. \figlabel{b} Maximum deterioration of the infidelity (only considering the contributions from noise) for model errors in $J_0$, $\eps_0$ and $\tau_{\mr{rise}}$ as large as 20\%.}
	\label{fig:bar}	
\end{figure}

With this setup it is then possible to calculate \IF as a function of \nseg pulses $\bveceps$ and stabilized \dbz, including decoherence from noise. We numerically search for gate implementations with minimal $\IF$ by using the Levenberg-Marquardt algorithm (LMA), which iteratively minimizes the Euclidean norm of a vector-valued objective function $\bvec{f}$ and features fast local convergence. Specifically, we solve the optimization problem
\begin{equation}
\min_{\bveceps} \left| \left(
\begin{array}{cccc}
 \IF_ {\dbz}(\bveceps), & \IF_ {\eps,\mr{slow}}(\bveceps), & \IF_ {\eps, \mr{fast}}(\bveceps), & \phi(\bveceps) \bvec{n}(\bveceps) - \phi_t \bvec{n_t}
\end{array} 
 \right) \right|^2,
\label{eq:opt}
\end{equation}
for fixed \ndbz and \nseg. We choose the first three vector components as the infidelity contributions of noise in \dbz, slow noise in \eps and fast noise in \eps. Additionally, we account for systematic deviations by adding the three components of $\phi(\bveceps) \bvec{n}(\bveceps) - \phi_t \bvec{n_t}$, where $\phi(\bveceps)$ and $\bvec{n}(\bveceps)$ describe the rotation angle and rotation axis of the gate realized in the current iteration. The subscript \textit{t} denotes the respective quantities for the target gate. These terms ensure that solutions have negligible contributions to $\IF$ from systematic errors, typically on the order of $\SI{e-10}{}$ or less. Furthermore, the minimization is subject to the previously detailed experimental constraints and bounds. In order to find a global optimum, we repeat the optimization 1000 times with randomly selected starting values. Sequences with low $\nseg$ are easier to implement experimentally, and high \dbz are unattractive because of increased relaxation during readout \cite{Barthel2012}. Thus, low $\nseg$ and \ndbz can cover the relevant search space.

\IF of the solutions for $\pi/2_x$-pulses is shown as a function of $\nseg$ and \ndbz in \rfig{fig:bar} (a), where $\IF < 0.7\%$ always. In the absence of noise, these gates give $U_t$ with insignificant systematic errors. The results for $\pi/2$-pulses around different axes orthogonal to the \dbz-axis, and for $\pi$-pulses, are qualitatively similar. We will therefore limit our discussion to $\pi/2_x$-pulses in the following. The best pulse with $\IF = \SI{1.1e-3}{}$ is found for $\ndbz = 3$ and $\nseg = 30$ (Fig. S.7 in \cite{supp}). The corresponding $\pi/2_y$-gate around the negative $y$-axis is slightly better with $\IF = \SI{0.9e-3}{}$ (Fig. S.8 in \cite{supp}). Typically, the main contribution to the infidelity comes from fast \eps-noise whose contribution to $\IF$ is generally larger by a factor of order unity than the contributions of slow charge and hyperfine noise. Therefore, the infidelities reported above improve to \SI{0.4e-3}{} if the noise model is not extrapolated as white but instead the $1/f^{0.7}$ decay is further extended to the \SI{}{GHz} range. 

Solutions with lower $\nseg$ have fewer degrees of freedom but also feature low \IF. Two $\pi/2$-gates around the \textit{x}- and negative \textit{y}-axis with $\nseg = 18$ and $\ndbz = 2$ with $\IF < \SI{2e-3}{}$ are shown in \rfig{fig:gates}. These gates are representative solutions, featuring distinct pulses in $J$ with the rest of the time spent at the baseline defined by $\eps_{\min}$. This avoids excursions to regions of high $J$ with a bigger sensitivity to charge noise $d J /d \eps \propto \exp(\eps/\eps_0)$. In between two $J$-pulses, the state vector rotates through approximately $2\pi$. Furthermore, pulse sequences around orthogonal axes (in the \textit{yz}-plane) are shifted by approximately $\pi/2$ with respect to each other \cite{supp}. Therefore, solutions can be interpreted as Rabi oscillations, which are corrected for experimental constraints, work without a rotating wave approximation and honor the constraint $J>0$ by excluding the negative half-waves.

It turns out that the solutions are partly decoupled from quasistatic charge and hyperfine noise. This can be seen from the first-order derivatives of $U(\bveceps,\dbz)$ w.r.t \eps and \dbz that are about an order of magnitude smaller than for simple $x$-rotations. Another indicator is the gates' filter functions \cite{Martinis2003, Cywinksi2008,Biercuk2009,Biercuk2011a} which peak at finite frequencies around \SI{100}{MHz}, similar to dynamical decoupling techniques like Hahn-echo or CPMG.

In an experiment it is likely that the functional form of \je and pulse edges will deviate from the ones used in the simulation. This will introduce systematic gate errors and also change the gates' sensitivity to noise. In panel (b) of \rfig{fig:bar} we consider only noise-related contributions to \IF and show that the noise properties of the gates are largely retained in spite of such deviations. $\IF_{\mr{worst}}$ denotes the worst outcome when $J_0$ and $\eps_0$ or $\tau_{\mr{rise}}$ are changed by $\pm \SI{20}{\%}$. For most gates, $\IF_{\mr{worst}}$ is still below 1\%, where simpler gates with fewer \nseg are usually better. The best result is found for $\nseg = 18, \ndbz = 2$ with $\IF_{\mr{worst}} = \SI{5.9e-3}{}$ as opposed to $\IF = \SI{1.5e-3}{}$. Our noise model should therefore reflect the experimental reality sufficiently if one aims for $\F \gtrsim 99\%$. However, systematic errors will contribute a few percent to \IF. 

\begin{table}[b]
\small
  \centering
  \caption{For small systematic gate errors, the measurement outcome $\mr{Tr}(\sigma_z U_i \Ket{S}\!\Bra{S} U_i^\dagger) = S_i$ depends linearly on the gates' axis and rotation angle errors \cite{Dobrovitski2010a}.}
  	\begin{ruledtabular}
    \begin{tabular}{rrcl}
    Sequences $U_i$ & Parameterization & & $S_i$ \\
    \hline
    $\pi/2_x$ & $-2\phi$  & = &$S_1$\\
    $\pi/2_y$ & $-2\chi$  & = &$S_2$\\
    $\pi/2_y \leftarrow \pi/2_x$ & $-n_y - n_z - v_x - v_z$  & = &$S_3$\\
    $\pi/2_x \leftarrow \pi/2_y$ & $-n_y + n_z - v_x + v_z$  & = &$S_4$\\
    $\pi/2_x \leftarrow \pi/2_x \leftarrow \pi/2_x \leftarrow \pi/2_y$ & $n_y + n_z + v_x - v_z$  & = &$S_5$\\
    $\pi/2_y \leftarrow \pi/2_x \leftarrow \pi/2_x \leftarrow \pi/2_x$ & $n_y - n_z + v_x + v_z$  & = &$S_6$\\
    \end{tabular}%
    \end{ruledtabular}
  \label{tab:bs}%
\end{table}%

In order to remove these errors we cannot rely on simulations, which inherently involve a potentially inaccurate model, but need to use actual experimental data. Quantum process tomography \cite{Nielsen2000, Chow2009} could be used to characterize a single gate's systematic errors but cannot be applied directly since only one readout axis $\sigma_z$ is naturally available via spin-to-charge conversion \cite{Petta2005}. Instead, one can self-consistently estimate the systematic errors of an entire set of gates using the bootstrap tomography method by \citet{Dobrovitski2010a}. This protocol is attractive not only because of its simplicity and self-consistency but also because it is first-order insensitive to decoherence for short gate durations. We hence propose, simulate and benchmark a self-consistent calibration routine which uses the bootstrap method for characterization and iterative removal of systematic gate errors. Our gate set contains both $\pi/2$-gates from \rfig{fig:gates} (around the $x$- and negative $y$-axis) and we measure $\sigma_z$ by projecting onto the ST-axis. If the gate sequences shown in \rtab{tab:bs} are each applied to the same initial state \qS, the measurement outcomes $S_i, i = 1 \ldots 6$ of each sequence depend on the gates' rotation-angle errors $\phi$ ($\chi$) and the axis-errors $n_y, n_z$ ($v_x, v_z$) of the $\pi/2_x$-gate ($\pi/2_y$-gate). Perfect gates give $S_i = 0$ and deviations are to lowest order linear in gate errors.

As before, we use the LMA to iteratively find gates with $S_i = 0$, i.e. solve $\min_{\bveceps, \pbveceps} \left| \bvec{S}(\bveceps,\pbveceps) \right|^2$, where $\bveceps$ and $\pbveceps$ denote the \eps-pulses of the $x$- and $y$-gate. Since only $\sigma_z$ is being measured, this protocol is invariant if both gates' rotation axes are jointly rotated around the $z$-axis. This does not pose a problem because we are only interested in obtaining an orthogonal gate set but one could introduce an additional measurement axis to circumvent this. However, solving this minimization problem would not lead to pulses with high fidelities since the gates' noise properties are not taken into account. We therefore add the infidelity due to noise $\IF_{n}$ of each gate to the optimization problem
\begin{equation}
\min_{\bveceps, \pbveceps} \left| \left(
\begin{array}{ccc}
 \bvec{S}(\bveceps, \pbveceps), & w_n \IF_{n}(\bveceps), & w_n' \IF_{n}(\pbveceps)
\end{array} 
 \right) \right|^2,
 \label{eq:bs_min_2}
\end{equation}
where $w_n, w_n'$ are heuristically chosen weights which take into account that the minimum of $\IF_{n}$ is generally different for both gates.

Measuring $\IF_{n}$ in an experiment is more involved than measuring $S_i$. Because we have shown before that the noise properties of the gates with few \eps-pulses are mostly unaffected even by large model errors, we choose instead to calculate $\IF_{n}$ theoretically in each iteration. Therefore, the algorithm is expected to remove systematic errors while largely retaining the gates' noise properties if $J_0$, $\eps_0$ and $\tau_{\mr{rise}}$ are known to sufficient accuracy.

\begin{figure}
\centering
\begin{picture}(8,3.5)
		\put(0,0){\includegraphics[width=8cm]{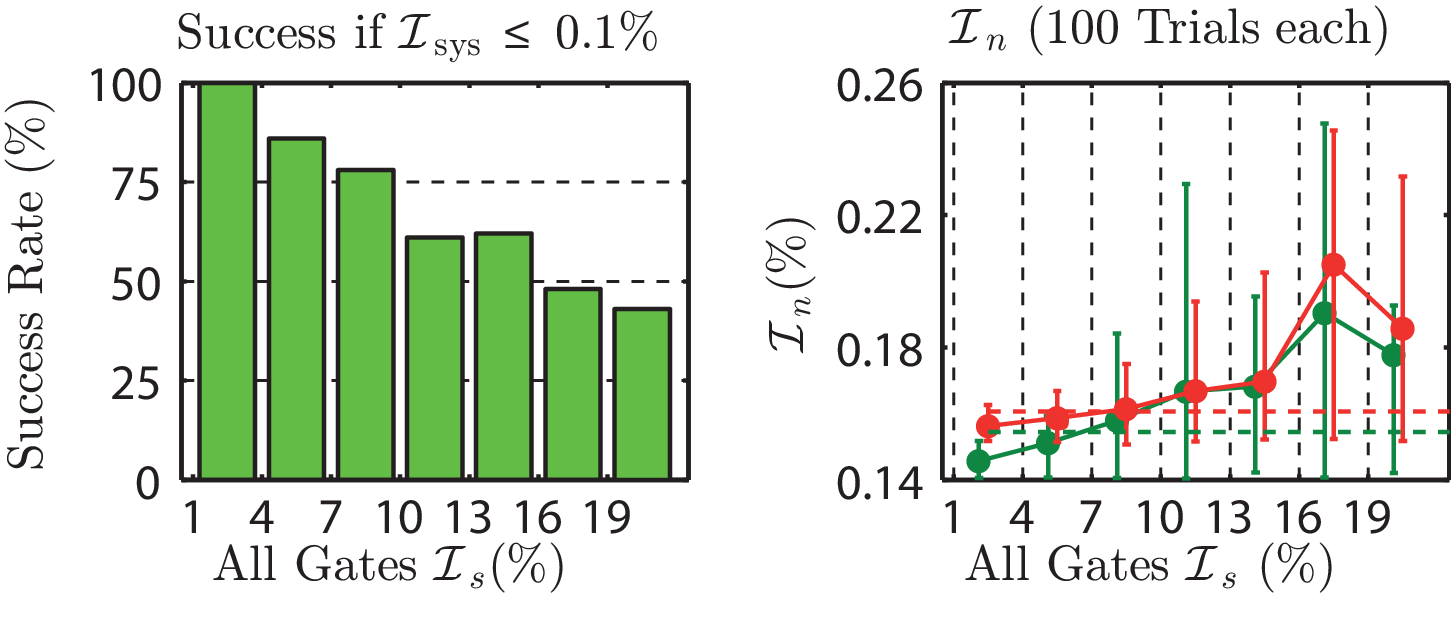}}			
		\put(-0.05,3.2){\figlabel{a}}	
		\put(4.1,3.2){\figlabel{b}}
  \end{picture}
\caption{(color online) \figlabel{a} The self-consistent tuning protocol converges even for bad initial infidelities $\IF_s$. \figlabel{b} The infidelity from noise $\IF_n$ of the final calibrated gates (dots) is on average close to $\IF_n$ of the perfect gates (dashed lines), and sometimes better because small systematic errors are now allowed. The error bars show the 10th and 90th percentile of the distribution of $\IF_n$ over 100 runs per bin for different starting gates.}
\label{fig:bs}
\end{figure}

We now benchmark the proposed calibration routine numerically. Randomly introducing systematic errors to the perfect gates found in the previous optimization (using the set from \rfig{fig:gates}), we find that our method converges for initial infidelities $\IF_s$ as high as \SI{20}{\%}, even when noise from averaging over a finite number (\SI{1e4}{}) of single shot measurements is taken into account (\rfig{fig:bs} (a)). The algorithm converges typically within 3 to 18 iterations where the exact rate depends on $\IF_s$ and \nseg of both gates. 
We call the algorithm successfully completed if the infidelity from systematic errors $\IF_{\mr{sys}}$ is smaller than $0.1\%$ for both gates \footnote{Since the calibration protocol is invariant under rotations around the measurement axis, we neglect the orientation of the rotation axes of both pulses in the \textit{xy}-plane. The calculation of $\IF_{\mr{sys}}$ and $\IF_s$ considers only the relative angle between both axes, rotation angle errors and deviations from the \textit{xy}-plane. See Supplemental Material for further details.}, but it usually reduces $\IF_{\mr{sys}}$ down to $10^{-4}$. Furthermore, the final gates are mostly as insensitive to noise as the perfect gates. As shown in \rfig{fig:bs} (b), better final results with lower $\IF_{n}$ are obtained if $\IF_s$ was small. Convergence within 10 iterations roughly corresponds to \SI{30}{min} in a current experimental setup (including measurement time and pulse updates on typical AWGs), which is realistic for experimental work.

In this work we have shown that high-fidelity single qubit gates exist for ST-qubits in GaAs. Based on measured noise characteristics we predict that the achievable fidelities are comparable to the thresholds of different QEC schemes. In order to eliminate systematic errors from these gates, we have developed and simulated a tuning algorithm based on experimental feedback. This algorithm works robustly in the presence of measurement noise and retains the gates' robustness to noise.

The results of this work will be used in the future to tune up a set of high-fidelity single-qubit gates, providing a valuable tool for performing accurate dynamical decoupling sequences, quantum state and process tomography. Furthermore, these gates will form the building blocks for two-qubit operations.

This work was supported by the Alfried Krupp von Bohlen und Halbach Foundation, DFG grant BL 1197/2-1 and the Alexander von Humboldt foundation. We would like to thank S. Mehl for many useful discussions.

Reprinted with permission from Pascal Cerfontaine, Tim Botzem, David P. DiVincenzo, and Hendrik Bluhm, Physical Review Letters 113, 150501 (2014). Copyright 2014 by the American Physical Society.

%
\end{document}